# Design space exploration of Ferroelectric FET based Processing-in-Memory DNN Accelerator

Insik Yoon, Matthew Jerry, *Student Member, IEEE,* Suman Datta, *Fellow*, *IEEE*, and Arijit Raychowdhury, *Senior Member*, *IEEE*

*Abstract*— *In this letter, we quantify the impact of device limitations on the classification accuracy of an artificial neural network, where the synaptic weights are implemented in a Ferroelectric FET (FeFET) based in-memory processing architecture. We explore a design-space consisting of the resolution of the analog-to-digital converter, number of bits per FeFET cell, and the neural network depth. We show how the system architecture, training models and over-parametrization can address some of the device limitations.*

## I. INTRODUCTION

As Deep Neural Network (DNN) based machine learning accelerators suffer from the memory bottleneck, research has started in earnest to develop alternative computing architectures that take advantage of *Processing-in-memory* (PIM) [2-10]. In particular, FeFET based non-volatile arrays, with low operating voltages, fast read/write cycles and ability to store multiple states in a single device are promising [2]; however, they suffer from several technology challenges. In particular, in spite of advanced pulse shaping, the conductance of FeFETs vary non-linearly with the number of pulses, which leads to a loss of classification accuracy when they are used as synaptic connections in a DNN. In this letter we quantify the effect of this non-linearity on classification accuracy and explore the design space with the FeFET and peripheral circuits to ascertain possible operating ranges and conditions. As opposed to earlier studies on the MNIST [2] data-set, we discuss the role on device imperfections on the classification accuracy of the more complex EMNIST data-set [1], which provides further insight into the design complexity for real-world applications.

## II. DEVICES AND ARCHITECTURE

### A. Device modeling

Fig 1 illustrates the measured polarization, transient current and amplitude modulated pulsing scheme for FeFETs, which results in an analog conductance change. For further details on the device fabrication, interested readers are pointed to [2]. Fig. 2(a) shows the model of FeFET conductance with respect to the number of write pulses validated against experimental data presented in [2]. We model the nonlinear conductance as

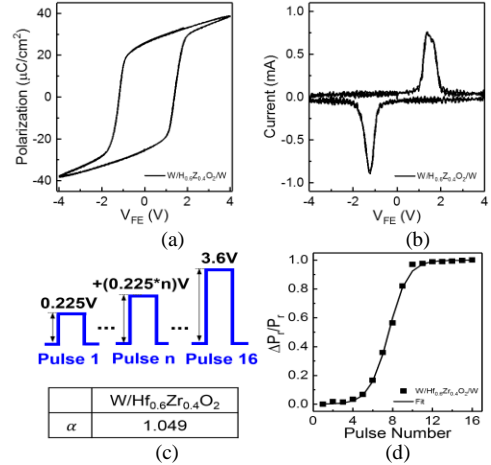

$$G(x) = \frac{\beta e^{\alpha x}}{1+e^{\alpha x}} + G_{\min}, \beta = G_{\max} - G_{\min} \quad (1)$$

where $G_{max}$ and $G_{min}$ are the maximum and minimum conductance values, and α is a parameter of nonlinearity (α=0.4 in Fig.2(a)). This is in contrast to the convex/concave functions that have been used in [2]-[4] to model non-linearity. We note that in the case of FeFETs the sigmoidal function is (1) a better fit and (2) physically meaningful. The sigmoidal conductance response manifests from the approximately Gaussian distribution of coercive fields among individual domains within the ferroelectric. Therefore, an amplitude modulated pulse scheme, which in essence, integrates across the domain distribution is expected to produce sigmoidal characteristics (Fig. 1). Varying α in Equation (1) controls the nonlinearity as shown in Fig. 2(b), and physically corresponds to changes in the distribution of the coercive fields within the H$_{0.5}$Z$_{0.5}$O$_2$ layer which is a strong function of the processing conditions including Zr concentration and the crystallization anneal temperature [16]. Fig. 2(b) illustrates the parametric space of FeFET conductance, α is varied from 0.25 to 2.0. The cell structure (Fig. 3) shows the schematic diagram for a differential FeFET memory cell capable of storing both positive and negative synaptic weights. We consider the default number of bits per FeFET cell is 5 (32 levels). The access transistors are connected to the gate of FeFETs, GL1 and GL2. During write, write pulses are applied through GL1 and GL2 when WL is asserted to high. During read, BL1 and BL2 are asserted to the read voltage and the currents from each row flowing through the SL are the summed via Kirchoff's circuit laws according to:

$$I_1 = -\Delta V(G-\Delta G), I_2 = \Delta V(G+\Delta G)$$
$$I = \Delta V(-G+\Delta G+G+\Delta G) = \Delta V(2\Delta G)$$

Fig. 1: Measured (a) polarization and (b) transient current as a function of $V_{FE}$ for a W/(10nm)Hf$_{0.4}$Zr$_{0.6}$O$_2$/W capacitor. (c) Amplitude modulated pulse scheme with pulse width=100ns directly samples the Gaussian domain distribution resulting in a sigmoidal response of the polarization to successive write pulses (d).

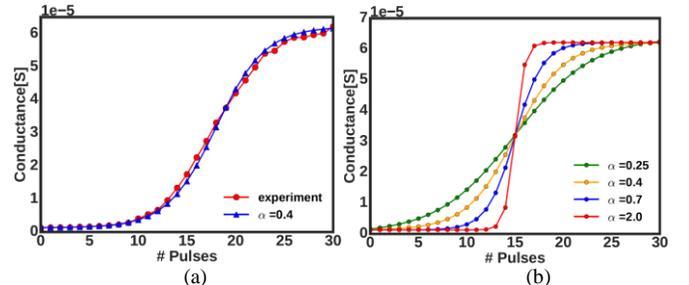

Fig. 2: (a) Comparison of the proposed conductance model with experimental data (α= 0.4) (b) Parametric conductance model for varying non-linearities

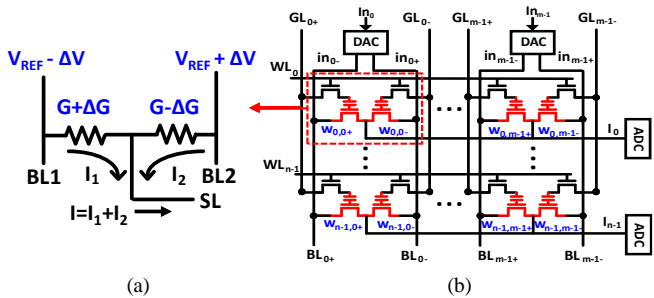

Fig.3: FeFET cell schematic (a) conceptual (b) PIM array architecture that enables (1 by m) input vector and (m by n) weight matrix multiplication. The red colored transistors are FeFETs which store the synaptic weights.

The differential cell structure allows us to generalize to both positive and negative numbers (both inputs and weights). As opposed to [4,12,13], this scheme (1) eliminates the need for using a separate array for storing negative weights and (2) enables the use of activation functions like "tanH"[17] and "leaky rectifier unit"[18] which are becoming more common in complex networks and whose outputs can be negative. Fig.3(b) shows the FeFET memory array for the multiplication of *1* by *m* vector as well as *m* by *n* matrices. SL of each cell in the same row are connected together. The vector inputs are processed by the per-column digital-to-analog converters (DAC). When all the WLs of FeFET array are activated simultaneously, the current sum at SL of each row represents the output of a matrix-vector multiplication. Therefore, by using such a PIM architecture, we can compute matrix-vector multiplication in *O(1)*[3]. The SL current is transferred to digital value after ADC and sigmoid operation is applied.

### B. Processing-In-Memory Architecture

The FeFET based PIM architecture for a DNN accelerator is shown in Fig.4, and is used to explore fully connected Deep Neural Network (DNN) and their performance on the EMNIST Balanced datasets [1] [14][15]. The network consists of 784 input nodes, a variable number of hidden neurons and 47 output neurons. Fig.4 shows the conceptual schematic of how the system operates. We envision distinct FeFET memory arrays which store the synaptic weights of the input to the hidden layer and 1, 2 or 3 hidden layers connecting to the output layer.

### C. Simulation Setup

We perform training and inference based on FeFET PIM architecture with EMNIST Balanced dataset on the Google Tensorflow platform augmented with device, circuit and architecture models to capture electrical non-idealities of the

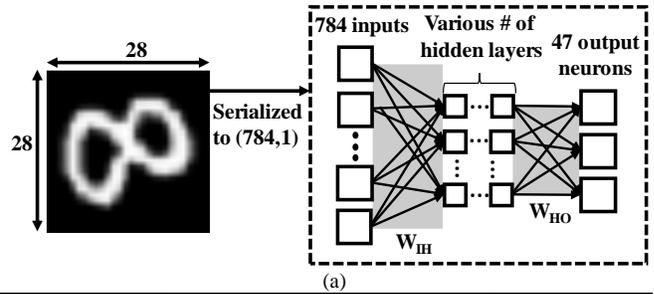

| # hidden layers in the network | # neurons in 1st layer | # neurons in 2nd layer | # neurons in 3rd layer |
|---|---|---|---|
| 1 | 100 | N/A | N/A |
| 2 | 256 | 100 | N/A |
| 3 | 484 | 256 | 100 |

(b)

Fig.4. (a)FeFET based PIM architecture for fully connected neural network and (b) Summary of the neural network architectures that are explored here.

hardware implementation. We compare three models. The first model is the **F**loating **P**oint(**FP**) model where training and inference are done on a GPU (baseline design). The second model is the **H**ardware-**A**ware **I**nference (**HAI**) model where inference is done on FeFET PIM architecture and the network training is hardware-agnostic and carried out on a GPU. During the training in HAI model, weights are scaled to the range [-1,1]. The third model is a **H**ardware-**A**ware **T**raining and **I**nference (**HATI**) model where both training and inference are done on the FeFET PIM. The classification accuracy from the FP model sets guidelines to assess inference performance of HAI model and training/inference performance of HATI model, as discussed in the following section.

### III. SYSTEM ANALYSIS AND RESULTS

Fig.5 illustrates how the classification accuracy of HAI model changes with the number of bits/cell of FeFET array and ADC bit resolution are used. The weights from FP model are quantized to 8 bit and stored in 8 bit/cell FeFET array when 8

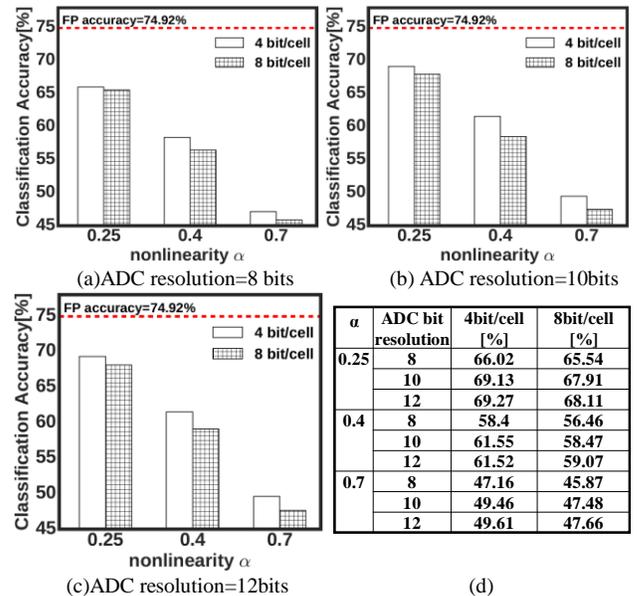

| α | ADC bit resolution | 4bit/cell [%] | 8bit/cell [%] |
|---|---|---|---|
| 0.25 | 8 | 66.02 | 65.54 |
|  | 10 | 69.13 | 67.91 |
|  | 12 | 69.27 | 68.11 |
| 0.4 | 8 | 58.4 | 56.46 |
|  | 10 | 61.55 | 58.47 |
|  | 12 | 61.52 | 59.07 |
| 0.7 | 8 | 47.16 | 45.87 |
|  | 10 | 49.46 | 47.48 |
|  | 12 | 49.61 | 47.66 |

Fig.5. (a)-(c) Classification accuracy for the HAI model with varying ADC resolution, non-linearity and the number of bits stored per FeFET cell. (d) Summary of the classification accuracy.

bit/cell FeFET array is used. In case of using 4 bit/cell FeFET

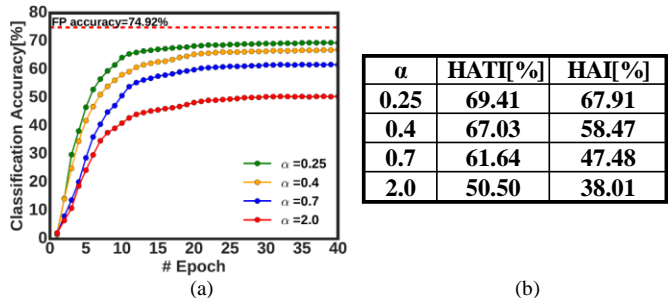

Fig.6 (a) Classification accuracy as a function of the nonlinear conductance of FeFET in HATI model. (b) Classification accuracy comparison between HATI and HAI models with respect to nonlinearity in conductance of FeFET.

array, 8bit quantized weights are split up into two 4bits and stored in 4 bits/cell FeFET array. Since we split the weights the area of FeFET array is doubled. From the figure, we observe that the classification accuracy with 4bit/cell FeFET array is better than the classification accuracy with 8bit/cell FeFET array. With fixed ADC bit resolution, degradation of precision in computation is worse when 8bit/cell FeFETs are used compared to a case when 4 bit/cell FeFETs are used. We also note that the accuracy increases as the number of ADC bits change from 10 to 12; but the increase is marginal. Therefore, with 784 cells/row in 8bits/cell FeFET array, 10bit ADC shows target performance under area/power/latency constraints. Previous investigations with simpler data-sets (MNIST) [2] have shown acceptable accuracy with the state-of-the-art FeFET devices. We note that more complex data-sets (EMNIST) impose higher restrictions on the device imperfections and further research needs to address them before technology adoption. Fig.6(a) shows how the classification accuracy changes with different nonlinear conductance curves from Fig.2(b) in HATI model and Fig.6(b) presents a comparison of the classification accuracy between the HATI and HAI models. From the figure, we observe that the classification accuracy drops by ~20% as α changes from 0.25 to 2.0. With the experimental data of FeFET conductance (α=0.4) shown in Fig.2(a), we achieve 67.38% of classification accuracy whereas the classification accuracy int eh FP model is 74.92%. This shows that even with hardware-aware training, a deep DNN model working on a complex data-set shows significant loss of accuracy because of device non-linearity. Fig.7 presents how classification accuracy changes with respect to the depth of the network. As the number of hidden layers increase, we observe that the classification accuracy drops in HAI model as shown in Fig7(a). The accuracy in HAI model drops because the errors in the matrix-vector multiplication propagate across the layers when the number of layers is large. However, HATI model shows an increase of accuracy as depth of network increases as shown in Fig.7(b). This is a key insight which shows that even if the devices themselves suffer from non-linearity, we can improve the classification accuracy by adding more hidden layers and making the model more complex (and hence, over-parametrized). This, of course, comes with an increase in training time and area/power; a discussion which is beyond the scope of this letter.

## IV. CONCLUSION

In this paper, we present how nonlinearity in conductance of FeFET significantly affects classification accuracy in training, necessitating the development of resistive memory technologies with higher linearity. We demonstrate that HATI models perform better that HAI models for higher non-linearity and the classification accuracy can be improved in HATI model by making the model over-parametrized.

## REFERENCES

[1] Cohen, G., Afshar, S., Tapson, J., & van Schaik, A. (2017). EMNIST: an extension of MNIST to handwritten letters.
[2] M. Jerry *et al.*, "Ferroelectric FET analog synapse for acceleration of deep neural network training," *2017 IEEE International Electron Devices Meeting (IEDM)*, San Francisco, CA, 2017, pp. 6.2.1-6.2.4. doi: 10.1109/IEDM.2017.8268338
[3] T. Gokmen, et al., "Acceleration of deep neural network training with resistive cross-point devices: Design considerations," Front. Neurosci., 10, 1–13, 2016
[4] S. Yu, et al., "Scaling-up resistive synaptic arrays for neuro-inspired architecture: Challenges and prospect," International Electron Devices Meeting, IEDM, 2016.
[5] L. Gao, et al., "Fully parallel write/read in resistive synaptic array for accelerating on-chip learning," Nanotechnology, 26, 45, 455204, 2015.
[6] S. H. Jo, et al., "Nanoscale memristor device as synapse in neuromorphic systems," Nano Lett., 10, 4, 1297–1301,2010
[7] S. Oh, et al., "HfZrOx -based Ferroelectric Synapse Device with 32 levels of Conductance States for Neuromorphic Applications," IEEE Electron Devices Lett., 99, 732–735, 2017
[8] S. Yu, et al., "Scaling-up resistive synaptic arrays for neuro-inspired architecture: Challenges and prospect," International Electron Devices Meeting, IEDM, 2016.
[9] R. Hasan, T. M. Taha, and C. Yakopcic, "On-chip training of memristor crossbar based multi-layer neural networks," Microelectron. J., vol. 66, pp. 31–40, 2017.
[10] M. Prezioso et al,. Training and operation of an integrated neuromorphic network based on metal-oxide memristors Nature, 521 (7550) (2015), pp. 61-64
[11] D. Soudry et al.,Memristor-based multilayer neural networks with online gradient descent training, IEEE Trans. Neural Netw. Learn. Syst. (99) (2015)
[12] M. Hu et al., "Memristor crossbar based hardware realization of BSB recall function," in Proc. Int. Joint Conf. Neural Networks, June 2012, pp. 1–7.
[13] B. Li et al., "Training itself: Mixed-signal training acceleration for memristor-based neural network,"in Proc. 19th Asia and South Pacific Design Automation Conf., 2014,pp. 361–366.
[14] Rosenblatt, Frank. Principles of Neurodynamics: Perceptrons and the Theory of Brain Mechanisms. Spartan Books, Washington DC, 1961
[15] Rumelhart, David E.; Hinton, Geoffrey E.; Williams, Ronald J. (8 October 1986). "Learning representations by back-propagating errors". Nature. 323 (6088): 533–536. doi:10.1038/323533a0.
[16] T. Schenk et al. About the deformation of ferroelectric hystereses Applied Physics Reviews 1, 041103 (2014); doi: 10.1063/1.4902396

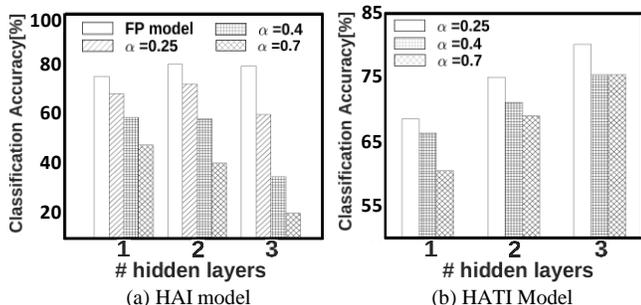

Fig. 7. Classification accuracy for FeFET array (8 bit/cell and 10bit ADC) as a function of the number of hidden layers.

[17] LeCun, Yann et al., "Efficient BackProp" Neural Networks: Tricks of the Trade, This Book is an Outgrowth of a 1996 NIPS Workshop, 1998, pages:9-50, isbn :3-540-65311-2
[18] Andrew L. Maas et al., "Rectifier nonlinearities improve neural network acoustic models" n ICML Workshop on Deep Learning for Audio, Speech and Language Processing, 2013